\documentclass[aps,prb,reprint,superscriptaddress]{revtex4-1}

\usepackage[T1]{fontenc}
\usepackage{graphicx}

\begin{document}

\title{Photoluminescence excitation and spectral hole burning spectroscopy of silicon vacancy centers in diamond}

\author{Carsten Arend}

\author{Jonas Nils Becker}
\affiliation{Fachrichtung 7.2 (Experimentalphysik), Universit\"at des Saarlandes,  Campus E2.6, 66123 Saarbr\"ucken, Germany}

\author{Hadwig Sternschulte}
\affiliation{Fakult\"at f\"ur Allgemeinwissenschaften, Hochschule Augsburg, An der Hochschule 1, 86161 Augsburg, Germany}
\affiliation{DiaCoating GmbH, 6112 Wattens, Austria}

\author{Doris Steinm\"uller-Nethl}
\affiliation{DiaCoating GmbH, 6112 Wattens, Austria}

\author{Christoph Becher}
\email[]{christoph.becher@physik.uni-saarland.de}
\affiliation{Fachrichtung 7.2 (Experimentalphysik), Universit\"at des Saarlandes,  Campus E2.6, 66123 Saarbr\"ucken, Germany}

\date{\today}

\begin{abstract}
Silicon-vacancy (SiV) centers in diamond are promising systems for quantum information applications due to their bright single photon emission and optically accessible spin states. Furthermore, SiV centers in low-strain diamond are insensitive to pertubations of the dielectric environment, i.e. they show very weak spectral diffusion. This property renders ensembles of SiV centers interesting for sensing applications. We here report on photoluminescence excitation (PLE) spectroscopy on an SiV ensemble in a low strain, CVD-grown high quality diamond layer, where we measure the fine structure with high resolution and obtain the linewidths and splittings of the SiV centers. We investigate the temperature dependence of the width and position of the fine structure peaks. Our measurements reveal linewidths of about 10 GHz as compared to a lifetime limited width on the order of 0.1 GHz. This difference arises from the inhomogeneous broadening of the transitions caused by residual strain. To overcome inhomogeneous broadening we use spectral hole burning spectroscopy which enables us to measure a nearly lifetime limited homogeneous linewidth of 279 MHz. Furthermore, we demonstrate evidence of coherent interaction in the system by driving a $\Lambda$-scheme. Additional measurements on single emitters created by ion implantation confirm the homogeneous linewidths seen in the spectral hole burning experiments and relate the ground state splitting to the decoherence rate.
\end{abstract}


\maketitle

\section{Introduction}
\begin{figure}
\includegraphics[width=1\columnwidth]{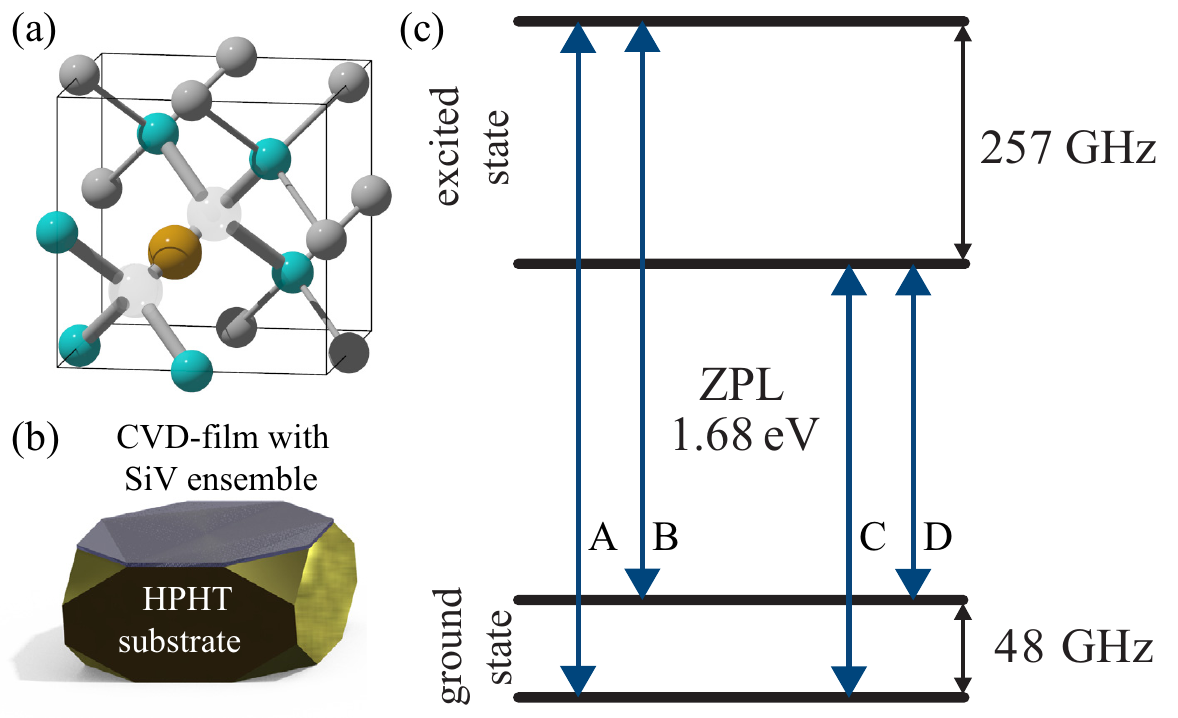}
\caption{(Color online) (a) Structure of the SiV center, consisting of a split vacancy (transparent) and an interstitial silicon atom (yellow) oriented along the [111] direction. Next-neighbor carbon atoms are shown in green. (b) The SiV ensemble sample is a commercial HPHT diamond overgrown with a thin homoepitaxial film in a CVD process. SiV centers are located in the thin film. (c) Electronic structure of the SiV center with split ground and excited states. The four possible ZPL transitions are denoted as A-D.
\label{fig1}}
\end{figure}

In recent years, color centers in diamond have attracted large attention for use in quantum communication and quantum information processing. Besides the well-known nitrogen vacancy (NV) center\cite{Clark1971,Acosta2013,Doherty2013}, the negatively charged silicon vacancy (SiV) center has emerged as a particularly interesting alternative. At room temperature, it provides narrow zero-phonon-lines (ZPLs) in the near infrared (738 nm) with line widths below 1 nm, weak phonon coupling (Huang-Rhys factor of 0.24) and count rates up to a few million counts per second.\cite{Neu2011} The SiV center is created by an interstitial silicon atom and a vacancy forming a split-vacancy structure along the $\langle$111$\rangle$ axes of the diamond lattice (see Fig. \ref{fig1}(a)), resulting in a doubly split ground and exited electronic state (see Fig. \ref{fig1}(c)).\cite{Goss1996,Hepp2014,Rogers2014c} This leads to a four line fine structure of the ZPL at cryogenic temperatures.\cite{Clark1995,Sternschulte1994,Sternschulte1995,Neu2013a,Rogers2014} Optical access to the fine structure of single SiV centers has been used to demonstrate spin-dependent fluorescence\cite{Muller2014} and coherent ground state spin manipulation.\cite{Pingault2014,Rogers2014b}

In this work, we aim at investigating optical access to the fine structure of large ensembles of SiV centers which is important for applications e.g. in high-sensitivity ensemble magnetometry\cite{Hong2013,Taylor2008}, collective nonlinear optical effects\cite{Harris1998} and optical quantum memories.\cite{Bussieres2013}
To this end, we use photoluminescence excitation (PLE) spectroscopy to examine a CVD-grown SiV ensemble. This allows us to measure the fine structure with high resolution and to analyse the temperature-dependent linewidth and line shift of the SiV centers. To overcome inhomogeneous broadening of the lines in the ensemble, we employ spectral hole burning spectroscopy which enables us to measure the homogeneous linewidth and compare it to the lifetime-limit. In order to compare ensemble results to data gained from single centers, we implant silicon ions into a high purity diamond which allows us to directly observe linewidths of single SiV centers in the absence of inhomogeneous broadening. One requirement for quantum information processing is the possibility to drive coherent processes in the system. By driving a $\Lambda$-scheme between two ground states and an excited state, we explore coherent optical interactions for the SiV center.

\section{Sample preparation and experimental setup}
The investigated SiV ensemble is contained in a single-crystal diamond layer which was grown on top of a (001) Ib high-pressure-high-temperature (HPHT) diamond substrate (Sumitomo) in a hot filament chemical vapor deposition (CVD) process (see Fig. \ref{fig1}(b)). The homoepitaxial growth prevents the formation of strain due to lattice or thermal expansion mismatch, thereby minimizing lineshifts and inhomogeneous broadening of the SiV center ZPL.\cite{Zaitsev2001} To facilitate high crystalline quality, the growth parameters were optimized, using a low methane fraction (0.26\% CH$_4$ in H$_2$) and a slow growth of approximately 10 nm h$^{-1}$, resulting in a layer thickness of 80-100 nm. Silicon contamination of the CVD chamber provides the silicon atoms which are incorporated into the diamond during the growth to form the SiV centers. For the second sample single SiV centers are produced by implantation of $^{28}$Si ions into a (111)-oriented type IIa HPHT diamond. After implantation, the sample is annealed for 3h at 1000$^\circ$C in vacuum and for 1h at 465$^\circ$C in air and cleaned in peroxomonosulfuric acid to remove graphite residues. To enhance fluorescence collection, solid immersion lenses (SILs) are fabricated on the surface of the diamond by focused ion beam milling (for further details, see Ref. \onlinecite{Pingault2014}).\\
The samples are mounted in a liquid helium flow cryostat (Janis Research, ST-500LN) integrated into a homebuilt confocal laser microscope. The temperature can be varied and stabilized between 5 K and room temperature by adjusting the helium flow and by electrical heating of the cold finger. A 100	x microscope objective with a numerical aperture of 0.8 (Olympus) focuses the laser light onto the sample and collects the fluorescence. The objective is placed inside the vacuum of the cryostat, eliminating the need for focusing through a cover glass and thereby considerably enhancing the collection efficiency. Excitation and fluorescence light are separated by a glass beam splitter and dielectric longpass and bandpass filters. After coupling into a multimode fiber, the fluorescence can either be measured with an avalanche photodiode (APD, Perkin Elmer SPCM-AQRH-14) to obtain the photon count rate or spectrally analysed with a grating spectrometer (Horiba Jobin Yvon, iHR 550 and Symphony CCD). A grating with 1800 grooves per mm provides a resolution of 0.06 nm (30 GHz) at 738 nm. The SiV centers are excited using a tunable (670-1000 nm) cw Ti:sapphire laser with a line width below 4 MHz (Sirah, Matisse) which enables non-resonant excitation at 700 nm and resonant excitation at 737 nm. The laser frequency can be scanned mode hop free over a range of 30 GHz. For spectral hole burning experiments, a homebuilt, tunable external cavity diode laser (ECDL) with a line width below 10 MHz is used as an additional excitation source. Both laser frequencies are monitored by a wavemeter with 50 MHz relative resolution (High Finesse, WS6/200). The ECDL frequency is actively stabilized to the wavemeter using a PID loop.\\
PLE spectroscopy is performed by scanning the Ti:sapphire laser over the fine structure transitions of the SiV ensemble. The resolution is then given by the resolution of the wavemeter and no longer limited by the spectrometer. The emitted photons are detected on the phonon sideband within a bandwidth of 760-800 nm, enabling suppression of the laser light with 750 nm longpass filters and preventing the need for complicated filtering schemes to detect the signal on the ZPL. Although the detected sideband intensity under non-resonant excitation is about two magnitudes weaker as compared to the ZPL, this detection method still gives high count rates due to the high efficiency of the resonant excitation.

\section{Results and discussion}
\subsection{Low temperature PLE spectrum}
\begin{figure}
\includegraphics[width=1\columnwidth]{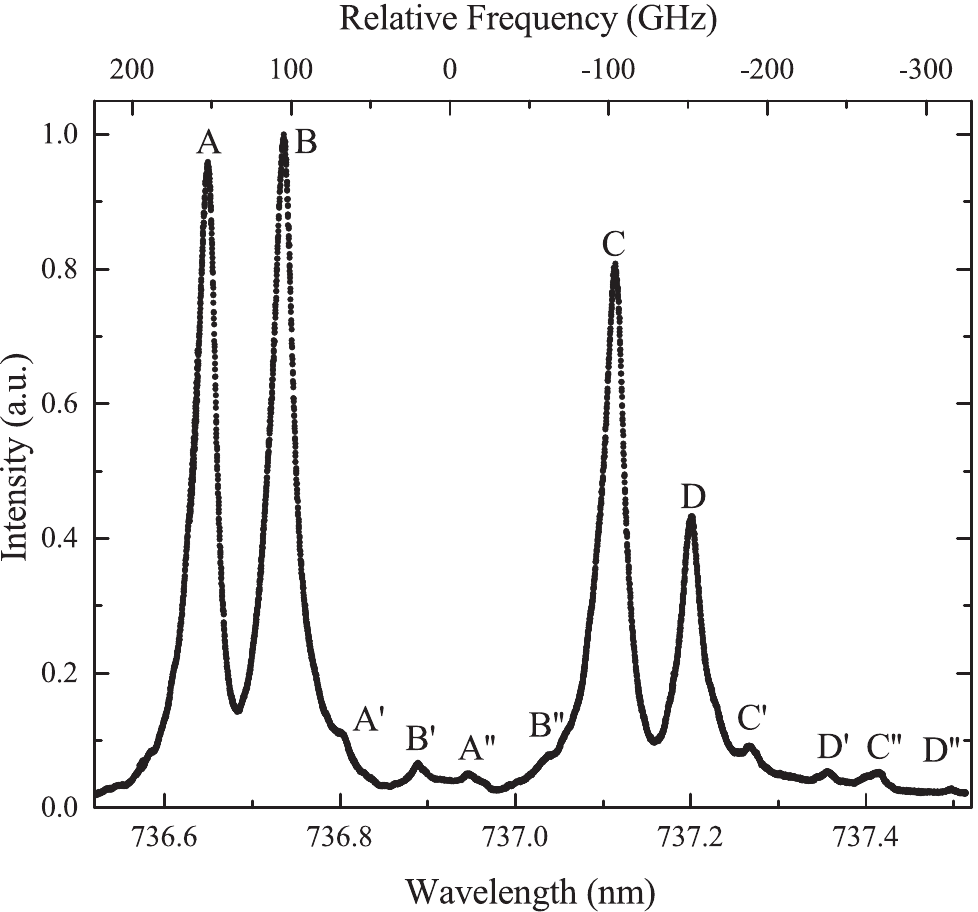}
\caption{PLE spectrum of the SiV ensemble at 5K. It is dominated by four strong peaks with individual linewidths of about 10 GHz arising from the electronic structure of the SiV center. Additional peaks are attributed to subensembles created by different silicon isotopes.\label{PLE5K}}
\end{figure}

We first measure the low temperature PLE spectrum of the SiV ensemble at 5 K, which is shown in Figure \ref{PLE5K}. We observe the characteristic fine structure of the SiV center with two strong doublets, which confirms the high crystal quality and low strain environment of our sample. The peaks have a linewidth of about 10 GHz and we determine the ground (excited) state splitting to be $48.3\pm 0.7$ GHz ($256.7\pm 0.8$ GHz) (see table \ref{splittings}). Comparing the intensities of the different subensembles visible in Figure \ref{PLE5K}, e.g. C:C$'$:C$''$, we find a ratio of 92.3:4.34:2.31, being nearly identical to the natural abundance of silicon isotopes $^{28}$Si:$^{29}$Si:$^{30}$Si of 92.2:4.7:3.1. This is clear evidence that the color center is created by silicon. All these results are in very good agreement with the literature.\cite{Clark1995,Sternschulte1994,Dietrich2014b} 

\begin{table}
\caption{Ground and excited state splittings of the SiV fine structure for the different silicon isotopes as seen in Figure \ref{PLE5K} \label{splittings}}
\begin{ruledtabular}
\begin{tabular}{cr|cr|cr}
$^{28}$Si & $\Delta\nu$ (GHz) & $^{29}$Si & $\Delta\nu$ (GHz) & $^{30}$Si & $\Delta\nu$ (GHz)\\
\hline A-B & 48.1 & A$'$-B$'$ & 47.1 & A$''$-B$''$ & 49.4 \\
C-D & 48.1 & C$'$-D$'$ & 48.4 & C$''$-D$''$ & 48.6 \\
A-C & 256.6 & A$'$-C$'$ & 255.1 & A$''$-C$''$ & 257.3 \\
B-D & 256.5 & B$'$-D$'$ & 257.2 & B$''$-D$''$ & 257.5 \\
\end{tabular}
\end{ruledtabular}
\end{table}

\subsection{Temperature-dependent measurements}
\begin{figure}
\includegraphics[width=1\columnwidth]{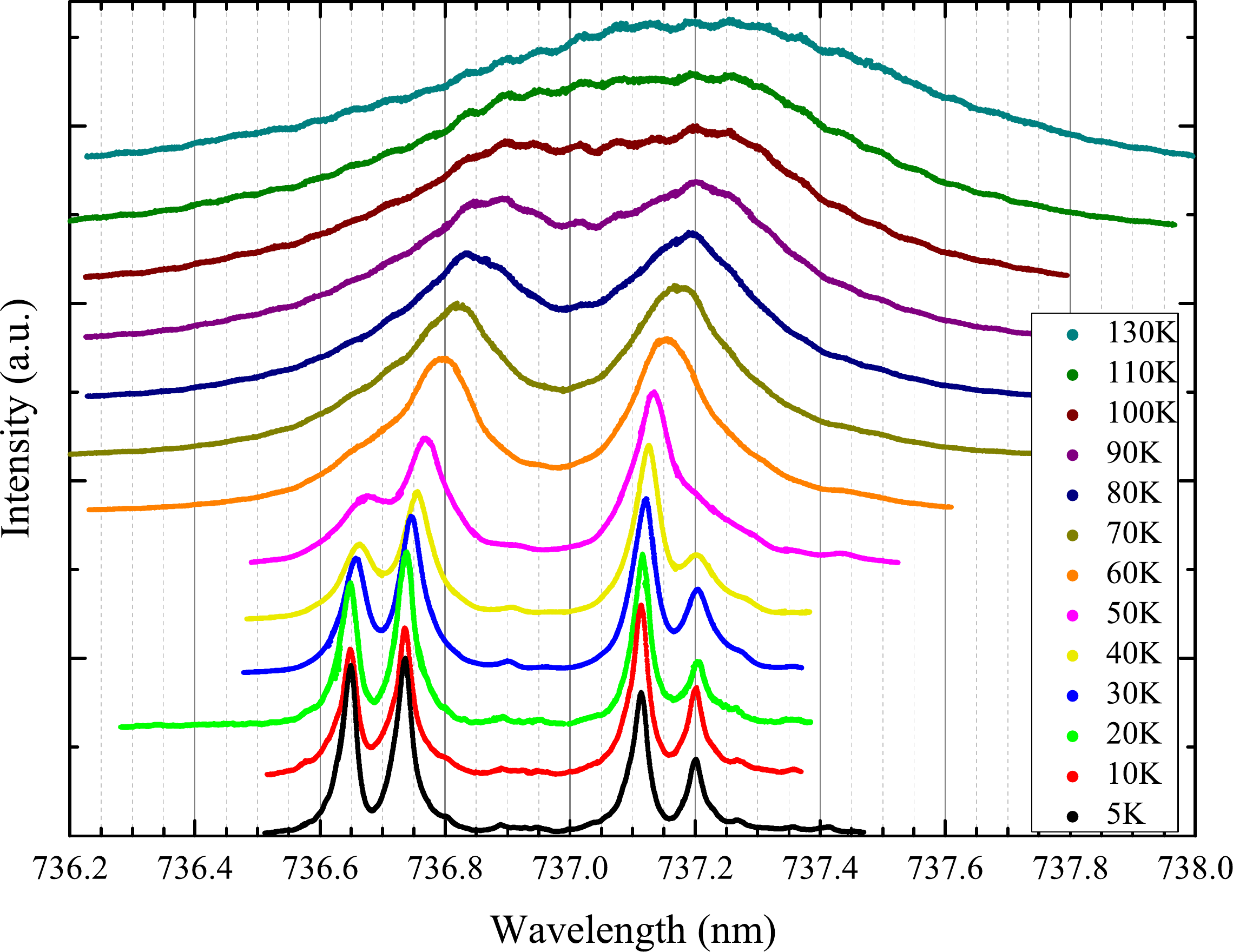}
\caption{(Color online) Temperature dependent PLE spectrum of the SiV ensemble. The four line fine structure only develops below 60 K. The peak positions are blue-shifted at lower temperatures. Oscillations in the spectra at higher temperatures are caused by residual laser power fluctuations in one mode hop free scanning range of 30 GHz. All spetra have been corrected for laser power fluctuations throughout the scans. \label{TempDep}}
\end{figure}

\begin{figure}
\includegraphics[width=1\columnwidth]{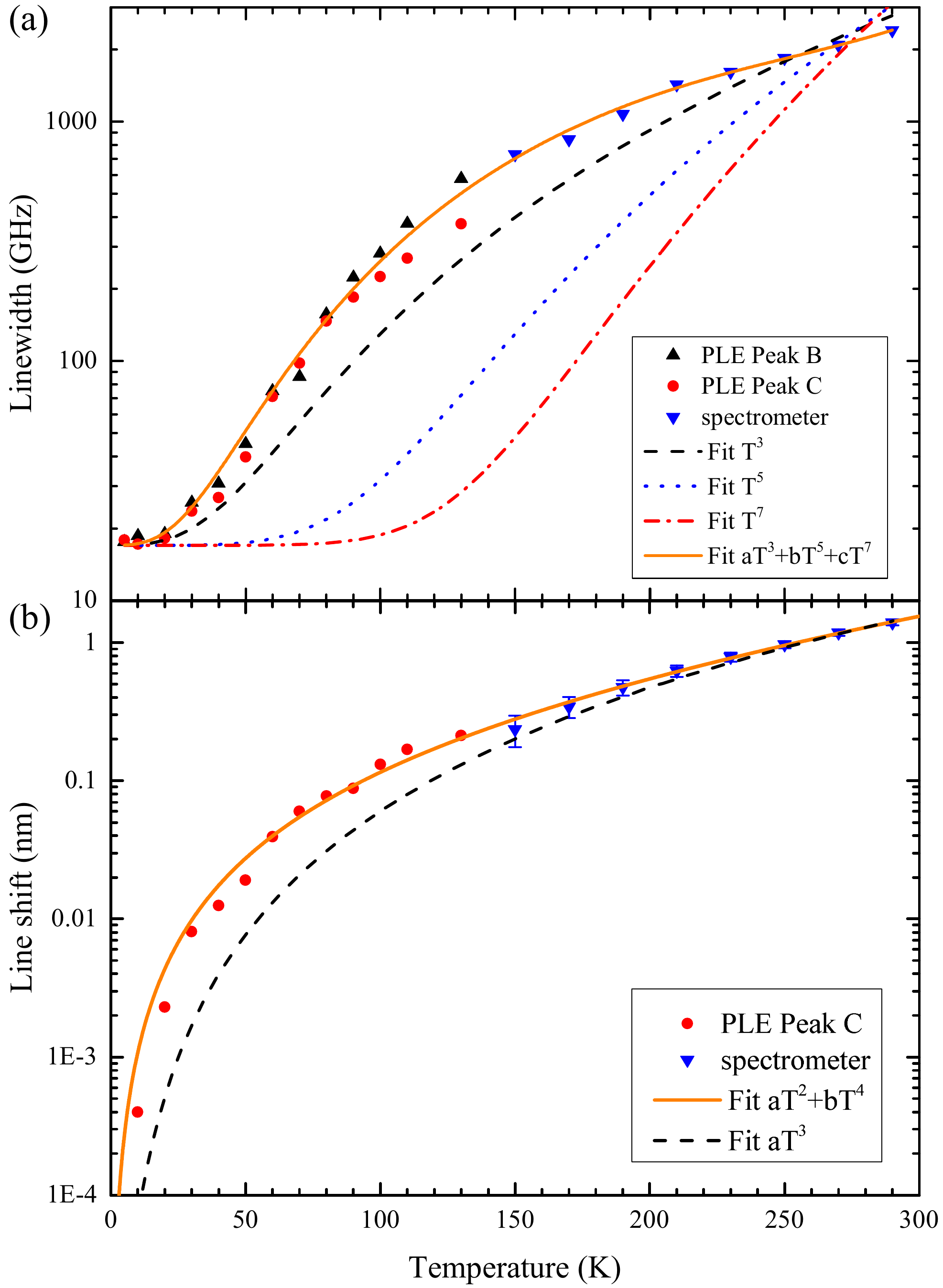}
\caption{(Color online) (a) Temperature-dependent linewidth of the SiV ensemble. At 130 K, peak B and C merge and the linewidth is measured using the spectrometer. Data from peak B is fitted using several dependencies on T. (b) Temperature-dependent line shift, analysis is performed using peak C.\label{width}}
\end{figure}

By varying the temperature from 5 K up to room temperature, we investigate the temperature dependence of the SiV fine structure spectrum. As displayed in Figure \ref{TempDep}, the fine structure peaks experience a broadening of the linewidth at higher temperatures. Above 60 K, the four line structure merges into two lines, and from 130 K on the fine structure is no longer visible. Additionally, the lines are blue-shifted more than 1 nm upon cooling down, as has been observed earlier using PLE\cite{Jahnke2014} or photoluminescence spectroscopy(PL).\cite{Neu2013a} The extracted linewidths for the main peaks B and C and the line shift for peak C are shown in Figure \ref{width}. For temperatures below 130 K, the spectra were fitted using a Voigt profile, showing both Gaussian and Lorentzian components. For temperatures above 130 K, measurements (as shown in Fig. \ref{width}) were performed using the spectrometer since the linewidths are well above its resolution limit and a Lorentzian profile was used to fit the data.

Both inhomogeneous and homogeneous line broadening are responsible for the observed linewidth. As the linewidth is nearly constant in the range of 5-20 K (see Fig. \ref{width}(a)), we assume that at low temperatures the broadening is dominated by temperature-independent inhomogeneous broadening due to residual strain in our sample. On the other hand, the homogeneous broadening is temperature-dependent and caused by quadratic electron phonon coupling. In reference to our earlier work (see Neu et al.\cite{Neu2013a}), where a detailed discussion of line broadening and shifting mechanisms is given, but a considerable effort had to be taken to account for the spectrometer resolution limit, we can now analyse the homogeneous line broadening using the high resolution data obtained by PLE spectroscopy. Several theoretical dependencies on the temperature $T$ have been proposed: Hizhnyakov et al.\cite{Hizhnyakov2002, Hizhnyakov2003, Hizhnyakov2004} derive a $aT^3+bT^7$ dependence in the presence of degenerate electronic states, verifying a dominant $T^3$ dependence for a NV center ensemble. They also deduce a $T^5$ dependence using perturbation theory in the presence of a dynamic Jahn-Teller effect. A $T^3$ dependence is attributed to phonon interactions with other defects in the crystal and is observed in the presence of strong inhomogeneous broadening.\cite{Hizhnyakov1999} Considering these possibilities, we fit our data with a $T^3$, $T^5$ and $T^7$ dependence (see Fig. \ref{width}(a)), taking into account the contribution of the constant inhomogeneous broadening. In good agreement with our earlier results, we find that the $T^3$ dependence indeed gives the best approximation to our data. Using a polynomial $aT^3+bT^5+cT^7$ fits the temperature-dependent linewidth very well and results in a dominant $T^3$ term. The concurrence of a dominant $T^3$ contribution to the temperature-dependent homogeneous broadening and a strong inhomogeneous broadening at low temperatures is in agreement with earlier theoretical predictions.\cite{Hizhnyakov1999} Recent measurements of single SiV centers in a high quality sample not showing inhomogeneous broadening also confirm the $T^3$ dependence for temperatures above 20 K.\cite{Jahnke2014} These measurements also show an additional contribution proportional to $T$ at low temperatures <20 K. As this contribution is small (<500 MHz) it is masked by the inhomogeneous broadening in our sample. The blue shift of the ZPL is a consequence of lattice contraction during cooling, leading to a temperature-dependent bandgap in diamond, and of quadratic electron-phonon coupling (for details, refer to Neu et al.\cite{Neu2013a}). Hizhnyakov et al.\cite{Hizhnyakov2002,Hizhnyakov2003,Hizhnyakov2004} derive a $aT^2+bT^4$ dependence, which we use to fit the measured line shift (see Fig. \ref{width}(b)) and which leads to a slightly better fit than the $T^3$ dependence recently observed by Jahnke et al.\cite{Jahnke2014} The high resolution of the PLE measurement enables us to confirm a good agreement of the fit with the data also for small shifts at low temperatures, a region not accessible to the spectrometer, and to support our earlier findings.

\subsection{Spectral hole burning in SiV ensemble}
For use in many quantum applications, one requires indistinguishable photons, i.e. fourier- or lifetime-limited photons. According to the equation\cite{Lounis2005} $\Delta\nu_{\text{lifetime}}=\frac{1}{2\pi\tau}$, the lifetime-limited linewidth is on the order of 100-150 MHZ for typical SiV center lifetimes of 1-1.7 ns. The lifetime of the investigated SiV ensemble is determined to be 1.6 ns at 5 K by using pulsed laser excitation and time correlated single photon counting, corresponding to a lifetime-limited linewidth of 100 MHz. This is in good agreement with 1.7 ns observed for single SiV centers at 4 K.\cite{Rogers2014}

\begin{figure}
\includegraphics[width=1\columnwidth]{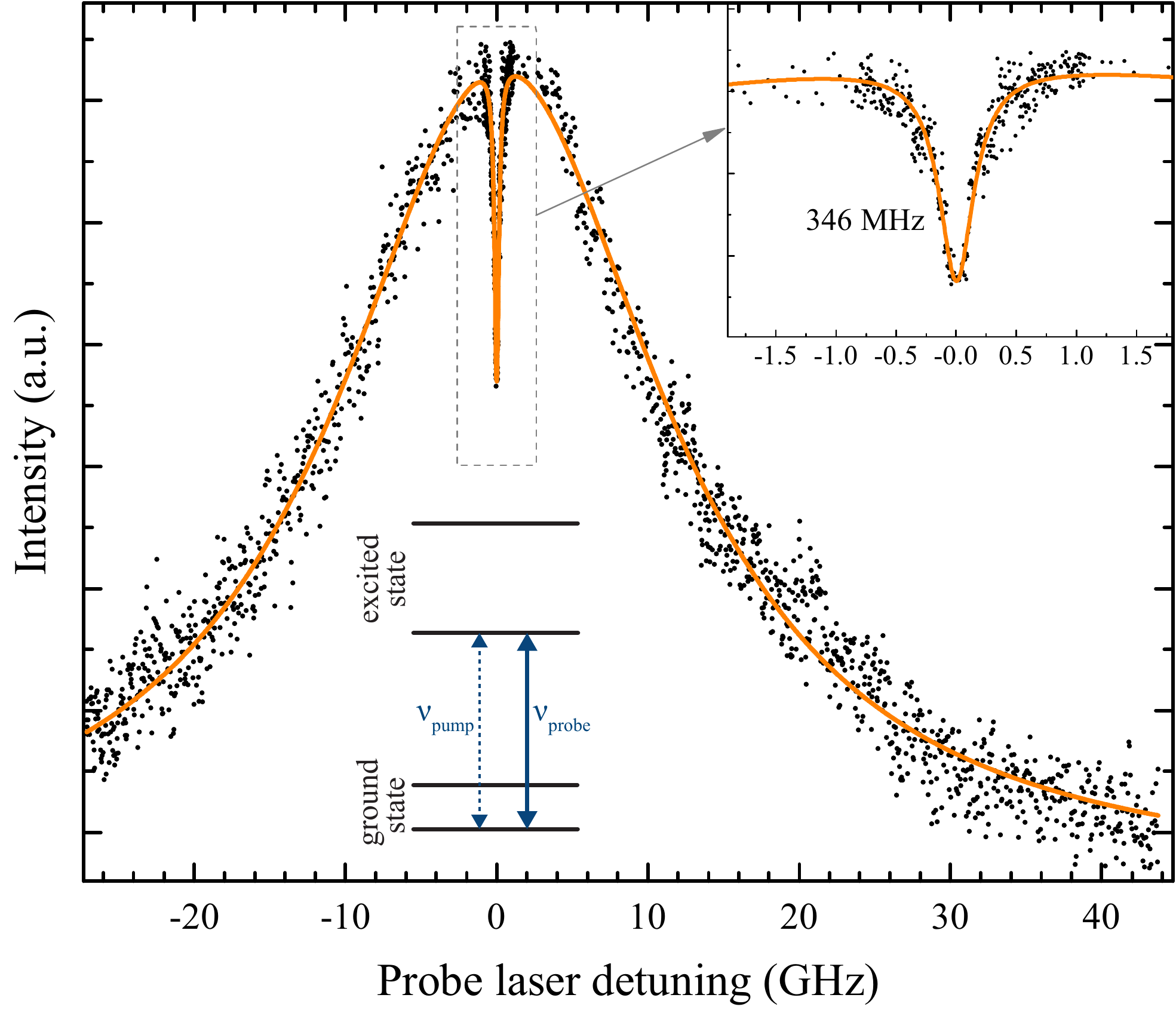}
\caption{(Color online) Spectral hole burning signal of transition C and corresponding level scheme. Probe laser frequency is shown in respect to the detuning from the pump laser frequency. The inset shows the power broadened spectral hole with a linewidth of 346 MHz. Data points are fitted using a bi-lorentzian function.\label{holeburning}}
\end{figure}

However, due to the inhomogeneous broadening, the measured linewidth of the SiV ensemble is two orders of magnitude larger than the lifetime limit. To determine the underlying homogeneous broadening, we employ spectral hole burning spectroscopy. The ECDL is used as pump laser and its frequency is fixed to the center of one of the fine structure transitions, thereby saturating a homogeneous subensemble of the inhomogeneous broadened transition. The Ti:sapphire laser is then scanned across the transition as weak probe laser. When both lasers are in resonance ($\nu_{\text{pump}}-\nu_{\text{probe}}=0$), we observe a sharp dip in the fluorescence signal, whose width corresponds to the homogeneous linewidth for low excitation powers. Figure \ref{holeburning} shows the spectral hole burning signal of peak C, featuring a hole with a linewidth of 346 MHz. This linewidth is subject to power broadening. By performing the measurement at different excitation powers, we extrapolate a linewidth of $279\pm16$ MHz for vanishing excitation power. Similar results are obtained for transition D. Witnessing these homogeneous linewidths close to the lifetime limit, we conclude that reaching the lifetime limit might be possible for single emitters in a very low strain sample (see also results in Sec. \ref{secD}).

\subsection{Single emitter spectroscopy}\label{secD}
\begin{figure*}[ht!]
\includegraphics[width=2\columnwidth]{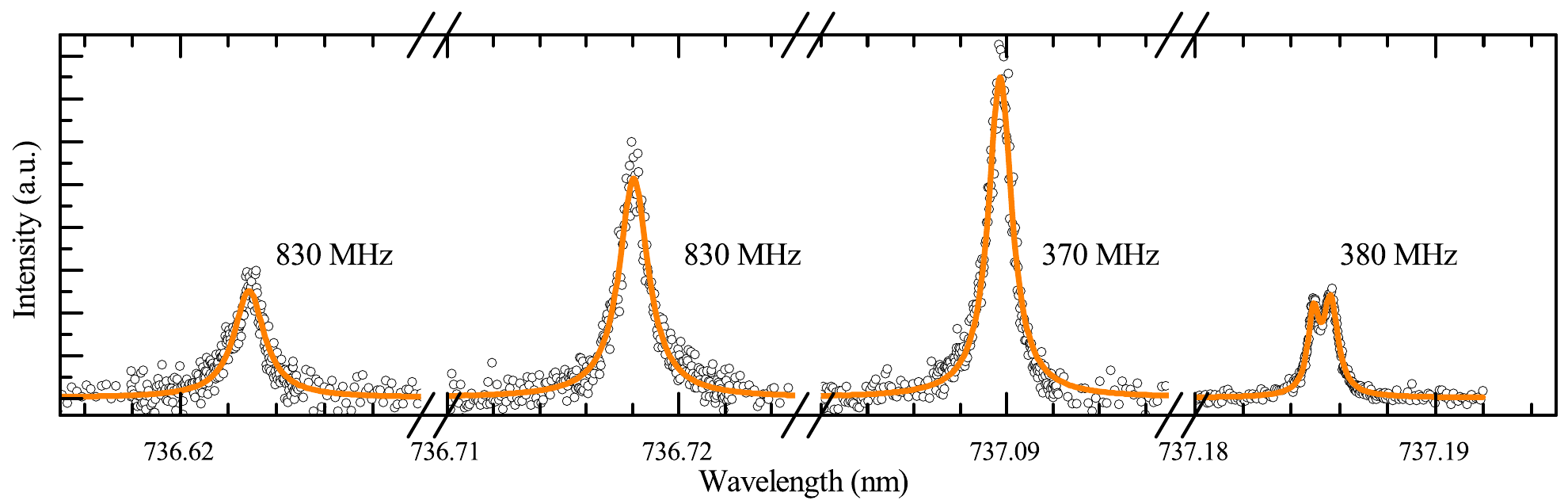}
\caption{(Color online) Power broadened PLE spectrum of a single SiV center in the SIL array. We extrapolate zero excitation power linewidths of 270 resp. 575 MHz, thereby showing linewidths close to the lifetime limit. We tentatively attribute the double peak structure of transition D to the presence of a second emitter in close spatial and spectral proximity.\label{ple_hpht}}
\end{figure*}

\begin{figure}[ht!]
\includegraphics[width=1\columnwidth]{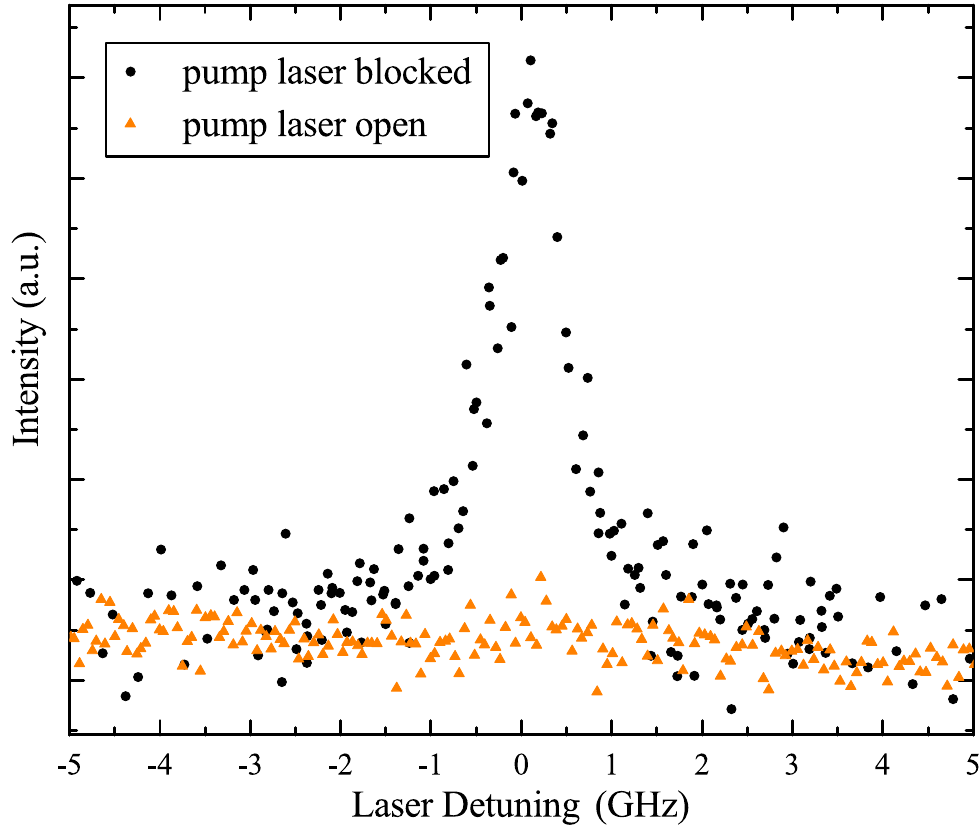}
\caption{(Color online) Spectral hole burning measurement of an emitter in the SIL array. When shining in the pump laser, the peak vanishes, indicating the absence of inhomogeneous broadening. \label{g2}}
\end{figure}

To test this assumption, we investigate the SIL arrays on the implanted sample and find bright emitters with count rates on the order of $10^5$ cps in photoluminescence. Measuring the intensity autocorrelation function g$^{(2)}(\tau)$, we obtain an anti-bunching dip below 0.5, proving single photon emission. Figure \ref{ple_hpht} shows the PLE spectrum of this emitter at 10K. Despite using very low excitation powers, power broadening of the lines is still present. We extrapolate the linewidths for vanishing excitation power and obtain values of 270 resp. 277 MHz for transitions C and D and of 575 MHz for transitions A and B. The results for transitions C and D are in good agreement with the spectral hole burning results found for the SiV ensemble and differ from the lifetime limit by only a factor of about 2. The larger linewidth of the transitions A and B is attributed to a shorter lifetime of the upper excited state due to fast decay into the lower excited state.\cite{Rogers2014} To verify the absence of inhomogeneous broadening, we again perform spectral hole burning measurements. As Figure \ref{g2} shows for transition C, the peak is completely saturated by the pump laser and therefore no longer visible when scanning with the probe laser. The lines can therefore only be homogeneously broadened. These results are confirmed by the recent demonstration of narrow linewidth single SiV centers by Rogers et al.\cite{Rogers2014}, where linewidths of 120-130 MHz have been measured for transitions C,D and 350-410 MHz for transitions A,B.

\begin{figure}[ht!]
\includegraphics[width=1\columnwidth]{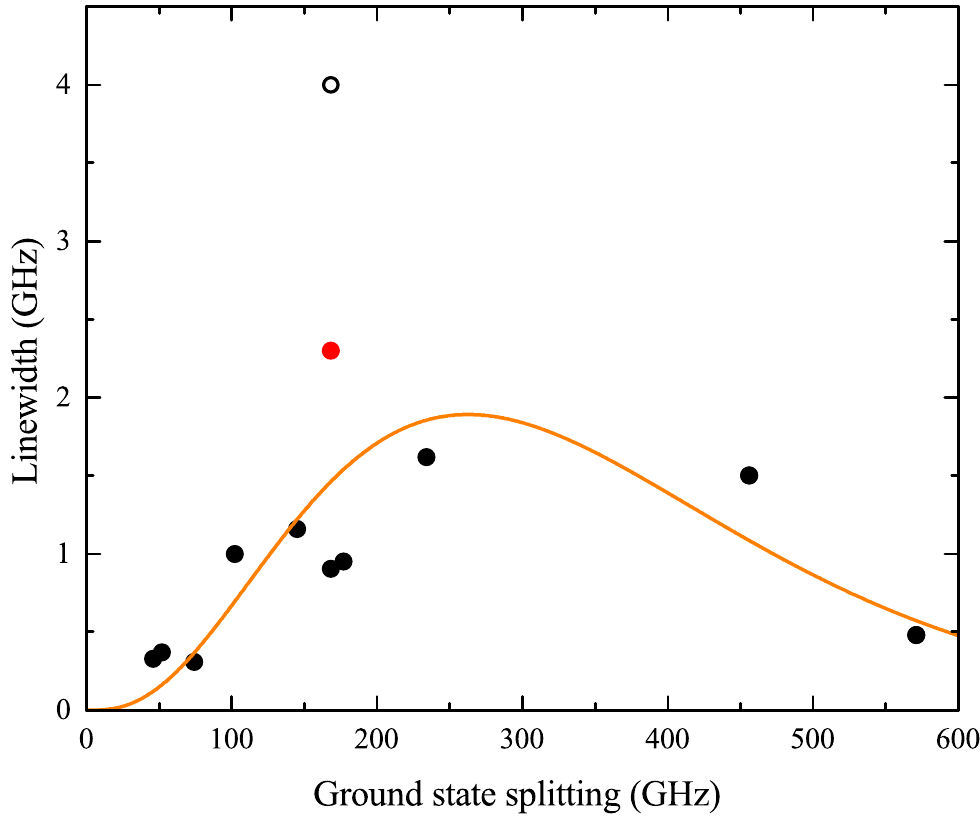}
\caption{(Color online) Linewidth of transition C as a function of the ground state splitting for twelve single SiV centers. The solid data points are fitted with the relation given in the text. The data point labeled with the open circle is omitted from the fit (see text).  The data point labeled with the red dot represents a strained single emitter for which a CPT measurement was performed (see Fig. \ref{cpt_strain}).\label{strain}}
\end{figure}

We assume that the slightly larger linewidths in our samples are due to residual strain. To verify this assumption, we investigate several single SiV centers and observe that the linewidth varies with the ground state splitting for different centers. To explain the observed variation, we use a model analogous to the model of Jahnke et. al.\cite{Jahnke2014}, where the linewidth $\gamma$ depends on the ground state splitting $\Delta$ and the level occupation probability in the following way: $\gamma=A\:2 \pi\:\Delta^3 (\exp(\hbar \Delta/k_B T)-1)^{-1}$, with $A$ as a free parameter (see Fig. \ref{strain}, solid data points and fitting curve). The change in linewidth can be related to a change in the decoherence rate within the ground and excited state manifold. For small splittings, the decoherence rate increases cubically with the phonon density of states and so does the linewidth. For larger splittings, however, the excitation probability drops exponentially with the increasing splitting. At around 275 GHz, the exponential part starts to dominate and therefore the linewidth peaks around this value. For even larger splittings, the linewidth decreases again due to the dominant exponential term. Such large state splittings can occur for emitters exhibiting a strong crystal strain which acts on the states similarly to the Jahn-Teller effect discussed in Ref. \onlinecite{Hepp2014}. A detailed discussion of the effects of crystal strain on the electronic properties of the SiV will be published elsewhere. We note, however, that not all data points are covered by this simple model (see data point labeled with open circle in Fig. \ref{strain}) indicating that the influence of strain on the electronic levels and their coherence might be more complex. In summary, we conclude that our implanted sample is an environment with larger strain as compared to the CVD-grown sample used by Rogers et al.\cite{Rogers2014}, which explains the slightly broader linewidths.

\begin{figure}[ht!]
\includegraphics[width=1\columnwidth]{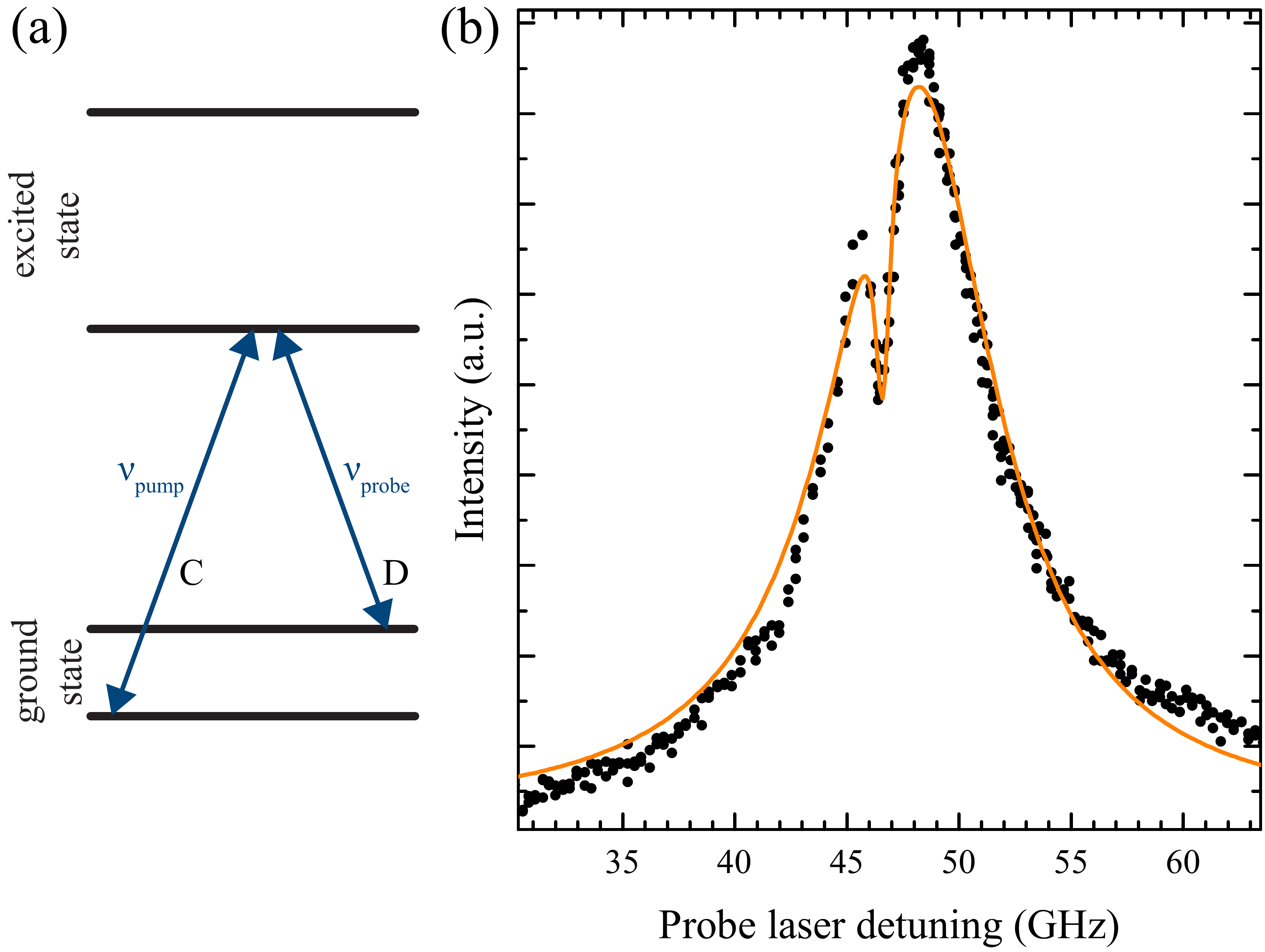}
\caption{(Color online) (a) $\Lambda$-scheme adressed by the two lasers on transitions C and D. (b) Fluorescence intensity of the SiV ensemble as a function of the two-laser-detuning. A dip is visible when both lasers are in resonance with the respective transition.\label{coherent}}
\end{figure}

\begin{figure}[ht!]
\includegraphics[width=1\columnwidth]{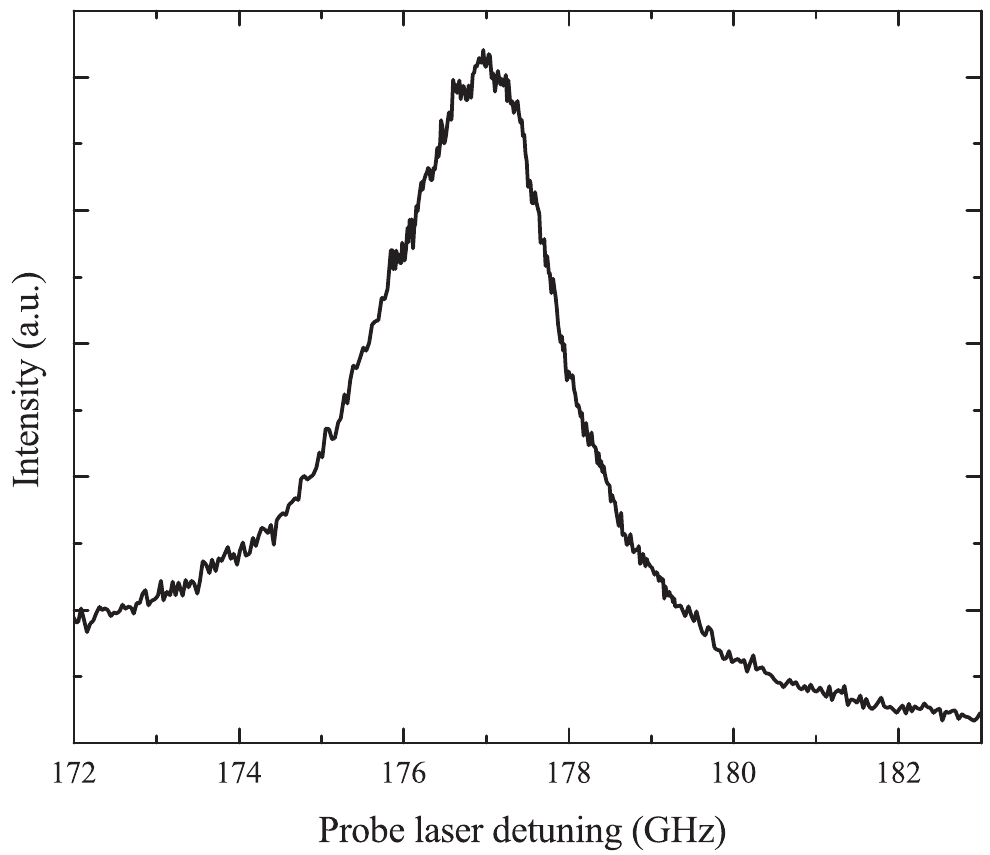}
\caption{CPT measurement for a single strained emitter (data point labeled with red dot in Fig. \ref{strain}). A $\Lambda$-scheme between transitions C and D is adressed. The CPT dip that is observed for emitters with smaller ground state splitting\cite{Pingault2014} completely vanishes due to the increased decoherence rate.\label{cpt_strain}}
\end{figure}

Another important prerequisite for quantum information processing is the availability of coherent interactions between electronic/spin states. By modifying our spectral hole burning experiments such that we fix the ECDL frequency on transition C and scan the second laser over the transition D, we can drive a $\Lambda$-scheme (Fig. \ref{coherent}(a)) in the SiV ensemble. As soon as the laser detuning equals the ground state splitting, a dip in the fluorescence intensity occurs, which is clear evidence of an coherent interaction, i.e. zero B-field coherent population trapping (CPT). A similar experiment for single SiV centers in the implanted sample recently demonstrated CPT between Zeeman-split levels in strong magnetic fields providing access to electron spin-coherence times.\cite{Pingault2014} With these results and the recently observed spin-conserving transitions for single SiV centers\cite{Muller2014}, all-optical access and control of orbital or spin degrees of freedom in single SiV centers seems feasible. However, our results highlight the importance of using SiV centers in a low strain environment for such experiments to avoid the detrimental influence of decoherence. To demonstrate this, we perform a CPT measurement for a strained SiV center (emitter labeled with red data point in Fig. \ref{strain}). As Figure \ref{cpt_strain} shows, no CPT dip is visible, which we attribute to the increased decoherence rate.

\section{conclusion}
In this work, we investigated the low temperature spectral properties of a high quality SiV center ensemble under resonant excitation. The high resolution of the PLE allows us to perform a detailed analysis of the temperature-dependent linewidth and line shift, especially for low temperatures. We demonstrate linewidths close to the lifetime-limit for an ensemble of SiV centers by eliminating inhomogeneous broadening using spectral hole burning and confirm these results by showing nearly lifetime-limited lines for single SiV centers. Furthermore, we linked the decoherence rate to strain in the sample by relating the ground state splitting to the linewidth and showed the negative effect of increased decoherence on CPT measurements. These results together with the observation of coherent processes prove the potential of both single SiV centers and SiV ensembles for applications in quantum information.

\begin{acknowledgments}
This research has been partially funded by the European Community's Seventh Framework Programme (FP7/2007-2013) under Grant Agreement No. 611143 (DIADEMS). We thank M.~Markham and A.M.~Edmonds of Element Six Limited for providing the HPHT (IIa) diamond sample. Ion implantation was performed at and supported by RUBION, the central unit of the Ruhr-Universit\"at Bochum. We thank D.~Rogalla for the implantation and C.~Pauly for the fabrication of solid immersion lenses. Moreover, we thank M.~Atat\"ure, B.~Pingault and C.~Hepp for helpful discussions throughout all stages of this work.
\end{acknowledgments}

\bibliography{library}

\end{document}